\documentclass[onetwocolumn,english,aps,superscriptaddress]{revtex4}

\usepackage{amstext}
\usepackage{graphicx}
\usepackage{amssymb}
\usepackage{setspace}
\usepackage{babel}
\usepackage{cases}
\usepackage{hyperref}
\begin{document}

\title{Robust optical delay lines via topological protection}

\author{Mohammad Hafezi \footnote{email: hafezi@umd.edu}}
\affiliation{Joint Quantum Institute, University of Maryland and NIST, College Park, MD 20742}
\author{Eugene E. Demler}
\affiliation{Physics Department, Harvard University, Cambridge, MA 02138}
\author{Mikhail D. Lukin}
\affiliation{Physics Department, Harvard University, Cambridge, MA 02138}
\author{Jacob M. Taylor}
\affiliation{Joint Quantum Institute, University of Maryland and NIST, College Park, MD 20742}
\setstretch{1.2}
\begin{abstract}

Phenomena associated with topological properties of physical systems are naturally robust against perturbations. This robustness is exemplified by quantized conductance and edge state transport in the quantum Hall and quantum spin Hall effects. Here we show how exploiting topological properties of optical systems can be used to implement robust photonic devices. We demonstrate how quantum spin Hall Hamiltonians can be created with linear optical elements using a network of coupled resonator optical waveguides (CROW) in two dimensions. We find that key features of quantum Hall systems, including the characteristic Hofstadter butterfly and robust edge state transport, can be obtained in such systems. As a specific application, we show that the topological protection can be used to dramatically improve the performance of optical delay lines and to overcome limitations related to disorder in  photonic technologies.

\end{abstract}

\maketitle

Particles in two-dimensional structures with a magnetic field exhibit
a remarkable variety of macroscopic quantum phenomena, including
integer \cite{Klitzing:1980} and fractional \cite{tsui82} quantum Hall
and quantum spin Hall effects \cite{Konig:2007}, and predicted regimes
of fractional or non-abelian statistics \cite{LesHouchesHall:BOOK,prange:BOOK}.  A hallmark of these systems is the presence of edge states, whose transport properties are robust against disorder and scattering, leading to quantized conductance sufficient to provide a resistance standard  \cite{Jeckelmann:2001,Novoselov:2007}.  Natural robustness of topological states is actively explored  in  quantum computation \cite{kitaev03,Nayak:2007}. Recently,
approaches to observing similar quantum Hall behavior in bosonic
systems including ultra-cold gases (for a review see
Ref.\cite{Cooper:2008p6565}) and photons \cite{Haldane:2008,Cho:2008,Wang:2008,Otterbach:2010,Koch:2010, Kitagawa:2010} have been suggested.

Our method for realization of topological protected photonic devices makes use of two dimensional arrays of coupled resonator optical waveguides (CROW) to simulate a 2D magnetic tight-binding Hamiltonian with degenerate clockwise and counter-clockwise modes. This approach does not require explicit time-reversal symmetry breaking \cite{Haldane:2008,Cho:2008,Wang:2008,Otterbach:2010,Koch:2010}, but the degenerate modes ---time-reversed pairs---
behave analogously to spins with spin-orbit coupling in 
the electronic quantum spin Hall effect (QSHE)
\cite{Haldane:1988,Kane1:2005,Bernevig:2006p27579}, and experience a
spin-dependent magnetic field (Fig.~\ref{fig:coupled_resonators}).
When the clockwise and counter-clockwise modes are decoupled, we can
selectively drive each mode and observe
quantum Hall behaviors without breaking the time-reversal symmetry in the
tight binding Hamiltonian. In a
direct analogy to the electronic integer quantum Hall effect, we show
that photonic edge states carry light at the perimeter of the system, while being insensitive to disorder,
and therefore forms a basis for robust photonic devices. In particular, in comparison to state-of-the-art 1-D CROW systems, our approach can be dramatically more resistant to scattering
disorders and fabrication errors.  

\section*{2D Photonic System and Quantum Spin Hall Hamiltonian}

\begin{center}

\begin{figure}
\includegraphics[width=0.90\textwidth]{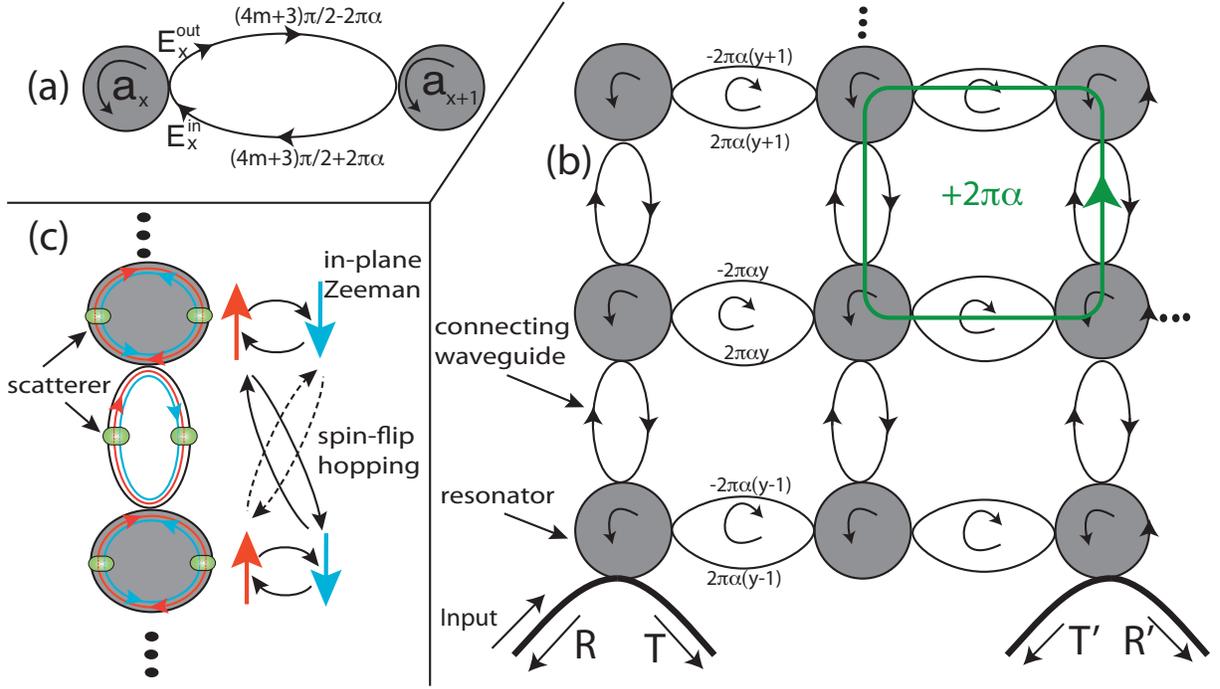}

\caption{\textbf{Schematics of the photonic system with a synthetic magnetic field: }(a) The dynamics of two coupled resonators is governed by a magnetic-type Hamiltonian (equation 3). The lengths of the upper and lower branch differ from each other so that the phase difference is $4\pi \alpha$ and m is an integer. (b) A magnetic tight-binding model can be implemented in a 2D lattice of coupled resonators. Only the waveguide phase differences are shown.  (c) By introducing scatterer in resonators and waveguides, one can induce in-plane magnetic field (Zeeman) and spin-flip hopping terms, in analogy to quantum spin Hall models.\label{fig:coupled_resonators}}

\end{figure}
\end{center}

As illustrated in Fig.~\ref{fig:coupled_resonators}, our system comprises optical ring microresonators that support
degenerate clockwise and counter-clockwise modes, restricted to one
pair per resonator.  We consider these modes as two components of a
pseudo-spin, i.e., clockwise ($\sigma=-1$, or psuedo-spin down) and
counter-clockwise ($\sigma=+1$, pseudo-spin up) circulation.
Resonators are evanescently coupled to each other and have been
studied in the context of 1D CROW \cite{Yariv:1999}, where the coupling leads to a tight-binding
model for photons and the corresponding photonic band structure.  By
coupling these modes in a two-dimensional arrangement, as we show below under appropriate
conditions, the dynamics of such photonic system is described by a
Hamiltonian for charged bosons on a square lattice (tight-binding),
but with the addition of a
perpendicular, spin-dependent effective magnetic field:

\begin{eqnarray}
  H_{0} & = & -\kappa \Big( \sum_{\sigma,x,y}\hat{a}_{\sigma x+1,y}^{\dagger}\hat{a}_{\sigma x,y}e^{-i2\pi\alpha y \sigma }+\hat{a}_{\sigma x,y}^{\dagger}\hat{a}_{\sigma x+1,y}e^{i2\pi\alpha y \sigma }\nonumber \\
  & + & \hat{a}_{\sigma x,y+1}^{\dagger}\hat{a}_{\sigma x,y}+\hat{a}_{\sigma x,y+1}^{\dagger}\hat{a}_{\sigma x,y} \Big) \label{eq:magnetic_tight_binding}\end{eqnarray}
where $\kappa$ is the tunneling rate of optical modes and $\hat{a}_{\sigma x,y}^{\dagger}$ is the
photon creation operator at resonator at site (x,y) with different
spin componenets ($\sigma=\pm1$). Specifically, photons acquire a
$2\pi\alpha \sigma$ phase when they go around a plaquette--equivalent
to having $\alpha$ quanta of magnetic flux penetrating each unit
plaquette \cite{Langbein:1969,hofstadter}.  


First, we show how the description of our system using such a Hamiltonian is derived.  We
start by considering two coupled resonators
(Fig.~\ref{fig:coupled_resonators}(a)), focusing only on the
counter-clockwise modes inside each resonator.  The length of
connecting waveguides is chosen such that photons destructively
(constructively) interfere inside the waveguide loop (resonator),
respectively, and therefore, they will be confined in the resonators
rather than waveguides.  Moreover, the lengths of the upper and lower
branches of the waveguide differ from each other, so when a photon hops from the left
to the right resonator, it acquires a  different phase than when it
hops in the opposite direction. This can be formally verified using
the standard input-output formalism \cite{Gardiner:1985}. In
particular, the boundary condition at the left resonator can be
written as
\begin{equation}
\hat{E}^{out}_x=\hat{E}^{in}_x+\sqrt{2\kappa}\hat{a}_x,
\end{equation} 
where $\hat{E}_x$'s are waveguide field operators at the vicinity of
the $x$-th resonator and $\hat{a}_x$ is the resonator field operator,
as shown in Fig.~\ref{fig:coupled_resonators}(a).  The resonator field
equation of motion is:
$\partial_t\hat{a}_x=-\kappa\hat{a}_x-\sqrt{2\kappa}\hat{E}^{in}_x$,
and similarly for the right resonator. Photons propagate freely
between the resonators, so for the upper branch we have
$E_{x+1}^{in}=-iE_x^{out} \exp({-2\pi i \alpha})$, and similarly for
the lower branch. By eliminating the waveguide fields, the left and
right resonator fields dynamics will be given by $\partial_t
\hat{a}_{x(x+1)}=i\kappa \exp{(\pm 2\pi i\alpha)}
\hat{a}_{x+1(x)}$, consistent with tunneling between resonators. Therefore, the effective Hamiltonian of the two
resonators takes the form:
\begin{equation}
H_{two-res}=-\kappa \hat{a}_{x+1}^\dagger\hat{a}_x e^{-2\pi i\alpha}-\kappa \hat{a}_{x}^\dagger\hat{a}_{x+1} e^{2\pi i\alpha}.
\end{equation} 

The above analysis for counter-clockwise modes (pseudo-spin up $\hat{a}_{\uparrow x,y}$) in the resonators shows that, in the absence of backscattering, they are decoupled from their time-reversed counterpart, i.e., the clockwise mode of the resonator (pseudo-spin down $\hat{a}_{\downarrow x,y}$).  At the same time, the pseudo-spin down component will experience a magnetic field similar to the pseudo-spin up component, where only the sign of magnetic field is changed ($\alpha \rightarrow -\alpha)$. Now,  by connecting resonators in a lattice structure and tuning the phase of the connecting waveguides, we can arrange  the acquired phase around each plaquette to be uniform and equal to $2\pi\alpha$. The phase can be tuned either by changing the length (or the index of refraction) of the connecting waveguides or by coupling ring resonators to the sides of the waveguides (similar to a Mach-Zender configuration \cite{Heebner:2004,Xia:2006}). The implementation of a Landau-type gauge is shown in Fig.~\ref{fig:coupled_resonators}(b) where the corresponding Hamiltonian is the form of equation (\ref{eq:magnetic_tight_binding}). Indeed, this is the trivial form of Hamiltonians with
QSHE in an analogy to time-reversal invariant
spin-orbit interactions in solid state
systems \cite{Haldane:1988,Kane1:2005,Bernevig:2006p27579}, where the magnetic field has
opposite orientations for the two pseudo-spin components.
Furthermore, one can control the coupling between different pseudo-spin component and exploit a wider class of QSHE Hamiltonians on a square lattice.  In particular,  semi-transparent scatterers inside the resonators or 
the connecting waveguides can be engineered to mix different
pseudo-spin components with each other. 

\section*{Generalized pseudo-spin-orbit interaction}

To illustrate this mixing, we consider the addition of a pair of
scatterers in every vertical connecting waveguide as shown in
Fig.~\ref{fig:coupled_resonators}(c). For simplicity, we assume the
scatterer is weak and loss-less. We characterize the strength of the
scatterer by a parameter $\epsilon$, where the transmission
coefficient is near unity $(t_{s}\simeq1)$ and the reflection
coefficient is $r_{s}=i\epsilon/\sqrt{2}$.  As shown in the
Supplementary Information (SI), the corresponding Hamiltonian of a
single vertical array of resonators will be: \begin{equation}
  H_{flip}=-\kappa\sum_{x,y}(\begin{array}{cc} \hat{a}_{\uparrow
      x,y+1}^{\dagger} & \hat{a}_{\downarrow
      x,y+1}^{\dagger}\end{array})\left(\begin{array}{cc}
      1 & \epsilon\\
      \epsilon & 1\end{array}\right)\left(\begin{array}{c}
      \hat{a}_{\uparrow x,y}\\
      \hat{a}_{\downarrow
        x,y}\end{array}\right)+h.c.\label{eq:spin-flip}\end{equation}
The diagonal terms are identical to the tight-biding terms of equation
(\ref{eq:magnetic_tight_binding}) and the off-diagonal terms represent
hopping between two adjacent sites while undergoing a spin-flip, i.e.,
the scatterers couple clockwise to counter-clockwise photons
(Fig.~\ref{fig:coupled_resonators}(c)). This spin-flip hopping is
similar to the Rashba term in the context of spin-orbit
interaction\cite{Rashba:1984,Kane1:2005}.

Similarly, if we consider a pair of weak scatterers inside the
resonators, then corresponding Hamiltonian takes the form:

\begin{eqnarray}
  H_{mag} & = & -\kappa\sum_{x,,y}(\begin{array}{cc}
    \hat{a}_{\uparrow x,y+1}^{\dagger} & \hat{a}_{\downarrow x,y+1}^{\dagger}\end{array})\left(\begin{array}{cc}
      1 & 0\\
      0 & 1\end{array}\right)\left(\begin{array}{c}
      \hat{a}_{\uparrow x,y}\\
      \hat{a}_{\downarrow x,y}\end{array}\right)+h.c. \nonumber \\
  & - & \frac{4\epsilon\kappa\mathcal{F}}{\pi}\sum_{x,y}(\begin{array}{cc}
    \hat{a}_{\uparrow x,y}^{\dagger} & \hat{a}_{\downarrow x,y}^{\dagger}\end{array})\left(\begin{array}{cc}
      0 & 1\\
      1& 0\end{array}\right)\left(\begin{array}{c}
      \hat{a}_{\uparrow x,y}\\
      \hat{a}_{\downarrow x,y}\end{array}\right).\label{eq:in-plane}\end{eqnarray}
The first term is the usual tight-binding form and the second term
represents the in-plane magnetic field which is enhanced by the finesse of the resonators (i.e., number of photon round trips $\mathcal{F}\simeq \pi/(1-r^2)$). If these vertical arrays replace the vertical
arrays of Fig.~\ref{fig:coupled_resonators}(b), then the overall
Hamiltonian of the system encompasses both an in-plane Zeeman term,
perpendicular to the synthetic magnetic  field (due to
on-site scatters) and a hopping-spin flip
term (similar to Rashba interaction).

\section*{Probing the photonic system}

We now show how optical spectroscopy measurements can be harnessed to analyze the transport properties of our photonic system.  In particular, as shown in Fig.~\ref{fig:coupled_resonators}(b), by
coupling two waveguides to the lattice edges, we evaluate the transmission
and the reflection of an input light field and study their properties
in connection with the magnetic states of the system (localized states and
edge states). 

We first consider a trivial case of quantum spin Hall effect where pseudo-spin-flip terms are absent. In this regime, we have two decoupled copies of regular quantum Hall states for opposite pseudo-spin components. We restrict our analysis to a single spin component,  and for brevity, we drop the spin index.  This choice enables us to examine in detail methods for probing the system and
determining its response to errors and disorders, without the additional complexity of spin Hall physics.

Using a formalism similar to the quantum scattering theory, we
investigate the problem of scattering of light field in optical
waveguides connected to our photonic system and evaluate transmission
and reflection coefficient under various conditions.  The waveguides only couple to
co-propagating modes in the resonators (counter-clockwise in
Fig.~\ref{fig:coupled_resonators}(b)), and thus under our assumption, the reflection in the input
channel and also transmission in the output channel is zero
(i.e., $R,T'=0$ shown in Fig.~\ref{fig:coupled_resonators}(b)).  The input-output probing waveguides are coupled to two resonators in the systems denoted by $|in\rangle$ and $|out\rangle$, respectively. As shown in SI, the self-energy of these resonators can be written as
 $\Sigma=-i\frac{\nu}{2}|in\rangle \langle in|-i\frac{\nu}{2}|out\rangle \langle out|$, where the coupling strength is
defined as $\nu$. Using Lippman-Schwinger equation, one can deduce different
reflection/transmission coefficients \cite{Fan:1999,Xu:2000}.  In
particular, the reflection coefficient is given by 

\begin{equation}
r'(\omega)=-i\nu\sum_{i\in in,j\in out} \left \langle i \left | \frac{1}{\omega-H_0-\Sigma} \right |  j \right \rangle\end{equation}
which means  an appreciable reflection should be observed when the frequency of an incoming photon becomes resonant with the energy of a photonic state inside the system. Note that if the photonic system is a single resonator, equation (6) reduces to the familiar form of: $r'(\omega)=\frac{-\nu}{\nu-i(\omega-\omega_0)}$.

\begin{figure}
\includegraphics[width=0.5\textwidth,height=0.40\textwidth]{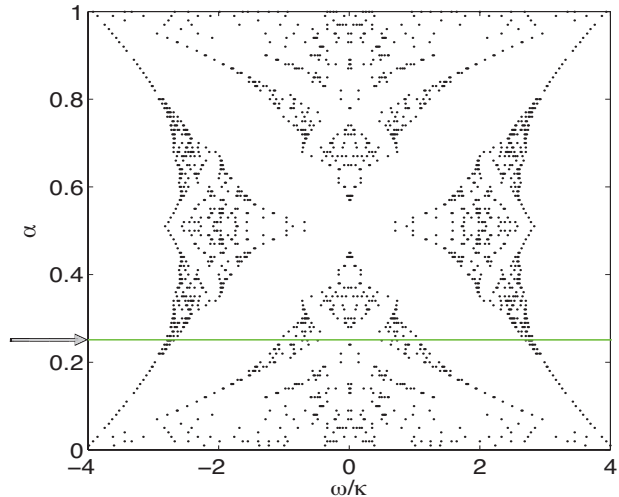}
\caption{\textbf{Hofstadter butterfly spectrum:} Each point represents a reflectivity greater than 0.005, for a 10x10 lattice with torus boundary condition and the coupling $\nu/\kappa=.02$. The green line shows the spectrum at the magnetic field of interest ($\alpha$) for the rest of the figures in the article.}
 \label{fig:Hofstadter-butterfly}

\end{figure}

The energy spectrum of the Hamiltonian of equation (\ref{eq:magnetic_tight_binding})
for an \emph{infinite} lattice is the well-known Hofstadter butterfly \cite{hofstadter}. We consider a $N_x \times N_y$ lattice with torus boundary condition
(i.e., coupling top-bottom and left-right edge together) to simulate
the effect of an infinite lattice. According to Hofstadter spectrum, for rational magnetic fluxes ($\alpha=p/q$), each magnetic band has many states ($\frac{N_{x}N_{y}}{q}$-fold),
which is the reminiscent of the Landau degeneracy in the continuum. The result of our numerical solution is shown in Fig.~\ref{fig:Hofstadter-butterfly} where the reflectivity $(R'=|r'(\omega)|^{2})$ is evaluated for different frequencies and magnetic field ($\alpha$), by the formalism described above. High reflectivity occurs when the
lower waveguide light is coupled to the system and completely transferred
to the reflection output channel (the second waveguide), similar to a channel drop filter.
We can readily see that the energy spectrum of the uncoupled system
(Hofstadter butterfly) can be obtained by measuring the system reflectivity.
We note that in order to to resolve different energy levels in the
spectrum, $\nu$ should be chosen to be sufficiently narrow ($\lesssim 8\kappa/(N_xN_y)$). 

\section*{Photonic edge states}

\begin{figure}[h]
\includegraphics[width=0.5\textwidth]{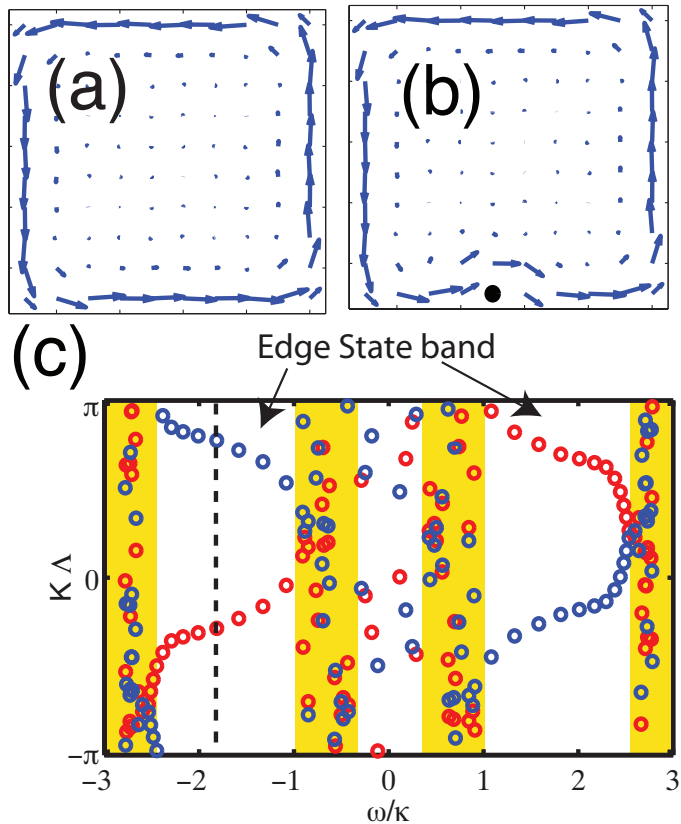}
\caption{\textbf{Edge states and their dispersion:} Probability current for an edge state in the absence
(a) and in the presence of a disorder (b). The location of the disorder is shown by the black dot, and $U/\kappa=5$. (c) shows the dispersion relation, a wave number is evaluated for each energy eigenstate at the lower (red) and upper edge (blue), i.e., $K\Lambda$ is the phase difference between two consecutive resonator at the edges. Magnetic bands state do not have well-defined wavenumber and are shown in the yellow area. The dashed line shows the state corresponding to (a).  In these plots, a 10x10 square lattice with $\alpha=\frac{1}{4}$ is considered. \label{fig:prob-current:pure/impure}}

\end{figure}

In a direct analogy to 2D electrons
in a magnetic field (quantum Hall physics) \cite{Halperin:1982,Rammal:1983,Hatsugai:1993PRB},
we recognize quasi-one dimensional states that are localized at the
perimeter of the system which carry current and are immune to disorders in form of random potential.
In contrast to toroidal boundary conditions where only magnetic bands
are present, in a finite square lattice, there exist states between magnetic bands
which are known as ``edge states''. These edge states carry a chiral current around the system
perimeter, as shown in Fig.~\ref{fig:prob-current:pure/impure}(a).
For each edge state, there is a similar edge state with an opposite
chirality which has the same energy magnitude but with the opposite sign.

So far we have considered a perfect system. To illustrate the robustness of the system to disorder, we consider a resonator that is detuned from its neighbors provides a model for ``non-magnetic'' disorder. This disorder can be readily characterized by an on-site potential at that site ($U\hat{a}_{x,y}^\dagger \hat{a}_{x,y}$). The main source of such frequency mismatch is small variations of the waveguide widths and the perimeters of the microrings during the fabrication process. Such imperfections are a common problem in photonics and prevents coupling large number of resonators \cite{Barwicz:2006,Xia:2007}.

In electronic quantum Hall systems, the edge states are immune to disorder \cite{prange:BOOK}. We find that such robustness applies to our photonics system. In particular, when such a disorder is located `inside' the edge state, the edge state is obviously not affected. However, when the disorder is
located on the edge, the edge state routes around it, as shown in Fig.~\ref{fig:prob-current:pure/impure}(b). More precisely, scattering which would reverse the current is prevented due to the lack of additional states with appropriate energy and momentum.

\section*{Application to delay lines}

\begin{figure}
\includegraphics[width=.97\textwidth]{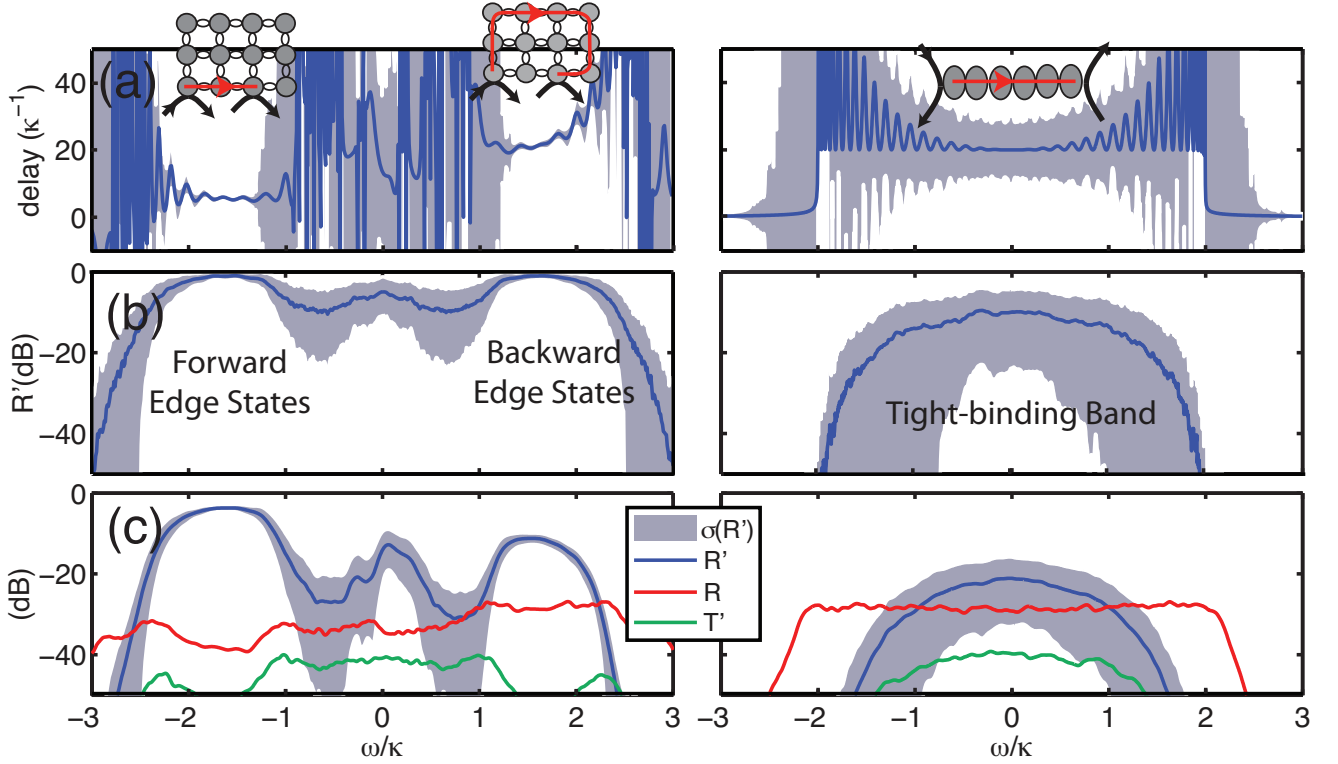}
\caption{\textbf{Edge states vs. CROW:}  This figure compares the quantum Hall system (left panels, 10x10 lattice) and CROW (right panels, array of 40) performances as delay lines. Blue curves show the average time delay (output reflectivity) in a(b), respectively, and the gray area highlights the standard deviations in the presence of  non-magnetic disorder (a gaussian disorder with a width $U/\kappa=0.4$ for 500 realizations). While the transport  is quite noisy in the magnetic bands and CROW (tight-binding band), the edge state bands exhibit noiseless transport with delays comparable to CROW. Depending on the input frequency, different edge states participate in the transport which leads to shorter or longer delays, as shown in the insets. In the presence of magnetic disorder  (the strength is a gaussian with a width $\epsilon\mathcal{F}=0.1$ and the phase is random $[0,2\pi)$) and loss ($\kappa_{in}=0.02\kappa$), the transport properties are more degraded for CROW than edge state bands.  While the counter-clockwise modes of the resonators is excited through an input field, the onsite scatterers backscatter photons in the clockwise modes. These modes leak out into $R$ and $T'$ channels which are non-zero in these plots. The coupling between input-output waveguides and the system is chosen to optimize the transport (for edge states $\nu=6 \kappa$ and for CROW $\nu=2\kappa$). In the quantum Hall system, the input and output waveguides are coupled to ($x=2,y=1$) and ($x=N_x-1,y=1$) resonators, respectively.\label{fig:transport}}

\end{figure}

The transport properties of edge states provide a robust alternative to conventional CROW in photonic delay lines.  In particular, we compare the transport properties of our photonic quantum Hall system to CROW, as illustrated in Fig.~\ref{fig:transport}.  In the quantum Hall system, there is a robust transfer band provided by edge states which carry photons from the input waveguide to the output waveguide (Fig.~\ref{fig:transport} inset), in a direct analogy to electronic edge state in the context of the integer quantum Hall effect \cite{Halperin:1982}.

In both systems, the operational bandwidth is given by smooth, linear part of the dispersion relation. In the quantum Hall system, the edge state band (Fig.~\ref{fig:prob-current:pure/impure}), is located between two Hofstadter bands (Fig.~\ref{fig:Hofstadter-butterfly}), while in the CROW configuration, the operational bandwidth is in the middle of the tight-binding dispersion to avoid the group velocity dispersion \cite{Yariv:1999}. Moreover, in both systems, the delay time is proportional to the number of resonators involved in the transport.  In the quantum Hall system, the transport can be either performed along the long or short edge of the system, depending on the input frequency, as shown schematically in Fig.~\ref{fig:transport} inset. Therefore,  in both systems in the absence of disorder, the bandwidth-delay product increases by the length (perimeter) of the system. However, in the presence of disorder, CROW and edge state behave differently as the system size increases. In particular,  in 1D systems (e.g. CROW) the disorder leads to localization \cite{Anderson:1958,Kramer:1993,Mookherjea:2008} and therefore, the transmission is impeded. In contrast,  the transport of edge states is topologically protected against disorder \cite{TKNdN,prange:BOOK} and the transmission is not affected (as discussed above for Fig.~\ref{fig:prob-current:pure/impure}).

\begin{figure}
\includegraphics[width=.4\textwidth]{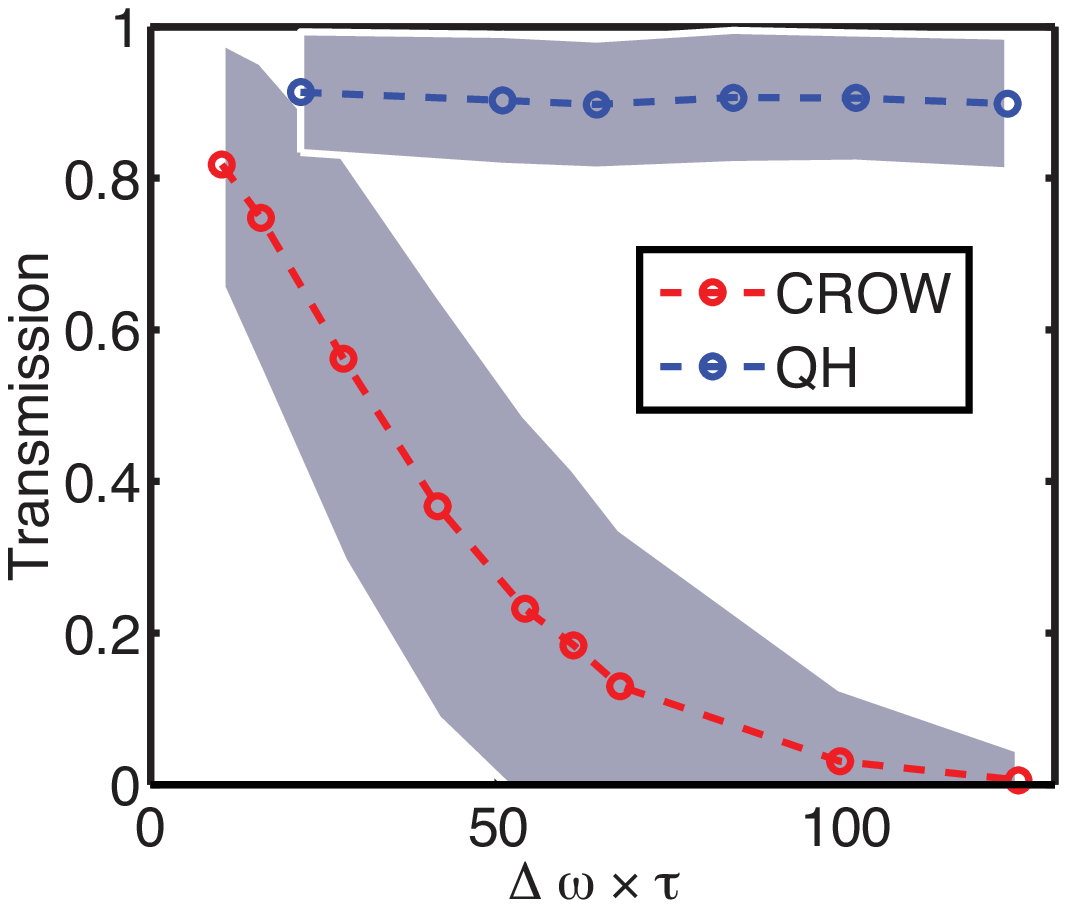}
\caption{\textbf{Transmission in 1D CROW and 2D quantum Hall system:}  For 1D CROW (2D quantum Hall), the transmission ($R'$) is evaluated at the center frequency $\omega=0$ ($\omega=1.5 \kappa$), where the group velocity dispersion is minimum, respectively. A gaussian disorder is assumed with a width $U/\kappa=0.4$ and the transmission is  averaged over an ensemble of 500 realizations. Gray shades show the standard deviation. By increasing the bandwidth-delay product (x-axis), the transmission for 1D CROW decreases  and the noise increases, while both the transmission and the noise in the 2D quantum Hall system remains constant.}  \label{fig:localization}

\end{figure}

To confirm that our edge states provide a robust transport, we numerically study the effect of disorder, by assigning a random frequency mismatch to the resonators. Taking average over many different disorder realizations, we observe that by increasing the size of the system and consequently the bandwidth-delay product, the transmission in CROW  decreases while the transmission with edge states is unaffected, as shown in Fig.~\ref{fig:localization}. Furthermore, we  observe that while magnetic band states and CROW depend sensitively on the disorder (position/strength), from one realization to another, the edge states are insensitive to the specific parameters, as shown in the standard deviation of the reflectivity and delay time in Fig. \ref{fig:transport}(a,b). The effect of the disorder on edge states is simply a shift of energies, which is manifested in a frequency shift in the edge states transfer band.   

We note that as time-reversal symmetry is not broken in our
system, we can not utilize such edge states as a one-way waveguide
similar to Ref.\cite{Wang:2009p16784}. More precisely, when we input
a light field in the backward direction into the system (by swapping
the input and output channel), the waveguides couple to the opposite
rotating field in the resonators (opposite pseudo-spin) and experience a magnetic field with
an opposite sign. Therefore, the system is reciprocal and the transport properties of the forward and backward feed are identical to each other.

Finally, we  investigate the effect
of loss in the resonators and other imperfections. Loss can be represented by a non-hermitian term in the Hamiltonian:
$-i\kappa_{in}\hat{a}_{x}^{\dagger}\hat{a}_{x}$ where the dissipation rate is given as the product of loss per roundtrip, number of roundtrips per decay time and the photon decay rate ($\kappa_{in}=\mu (\frac{\mathcal{F}}{2\pi}) (4\kappa)$) (see SI). The photonic loss can attenuate the reflection in the edge state transfer band due to the propagation around perimeter (Fig.~\ref{fig:transport}(c)). Silicon-on-insulator technology, where up to 100 micro-ring resonators have  been successfully coupled to
each other \cite{Xia:2007}, is a promising candidate for implementation of our scheme (see SI for experimental relevant parameters).

Other types of imperfection such as surface roughness can cause undesired backscattering which mixes  pseudo-spin up and down, acting as ``magnetic disorder''. As show in the SI,  these imperfections can be modeled by a magnetic disorder Hamiltonian. The backscattering effect manifests in the reduction of signal in $R',T$ channels and some leakage in $R,T'$ channels.  Fig.~\ref{fig:transport}(c) shows different transport coefficients. We observe that although the transport properties of magnetic bands states are affected by such magnetic disorder, edge state transport is robust, which is due the suppression of backscattering events. In particular, the scattering of a forward-going spin-up into forward (backward) -going spin-down is inhibited due to energy (momentum) mismatch.

\section*{Outlook}
Optical signals might be a promising alternative to electronic signals in the future circuits \cite{semi_roadmap:2010}. One of the key requirements is the ability to filter and slow down light on-chip over a large bandwidth (several Gbps)  for  various  time-domain processing such as optical buffering and multiplexing \cite{Baba:2008}. However, the effect of disorder in the millimeter size footprint is detrimental, e.g.  unwanted signal modulation in transmission spectrum \cite{Xia:2007}. Our system provides a platform to realize photonic system immune to disorder. Potentially, one can envisage using other types of topological properties to improve photonic technologies.

In addition, our photonic system enables a new approach for exploration of various fundamental quantum Hall phenomena.  This photonic system not only
enables investigations of quantum Hall
physics by simulating different types of Hamiltonians at room temperature, but it also taps into topological features to provide robust  devices for photonics. In the non-interacting regime (which was the topic of this article), one can
explore the Hofstadter butterfly of photons, and photonic edge
states as delay lines immune to disorders and also localization in 2D for non-interacting particles \cite{Huckestein:1995}. Furthermore, with the addition of interaction between photons, this system opens up exciting
prospects for exploring many-body topological state of light.  Although the ground state properties of such systems have been  extensively studied, the suitable characterization and measurement of strongly interacting photons is still an open question. In particular,  the chemical energy is absent for photons and the relevant conditions to study photons involves an externally driven system which leads to non-equilibrium situations \cite{Hafezi:2009a}. Another advantage of photons is their flexibility to form various system topologies (torus with different genera) by simply connecting waveguides to each other and manipulate such states for topological quantum computation. Intriguing additional applications of these ideas await further exploration.

\bibliographystyle{apsrev}

\section*{Acknowledgements}
The authors wish to thank Glenn Solomon, Edo Waks and Shanhui Fan for helpful discussions. This research was supported by the U.S. Army Research Office MURI award W911NF0910406, NSF,  AFOSR Quantum Simulation MURI, ARO-MURI on Atomtronics, Packard, DARPA QUEST, DARPA OLE and Harvard-MIT CUA.

\begin{center}
{\Large Supplementary Information}
\end{center}

\section{Transfer matrix formalism and tight-binding model}

Here we derive the magnetic tight-biding model which is presented in the main text as equation (1). In contrast to the toy model of two coupled resonators, presented in the main text, we consider an infinite array of coupled resonators similar to CROW \cite{Yariv:1999}. Our system consists of an array of coupled resonator and waveguides which is periodic and linear. Therefore, the transfer matrix formalism and Bloch's theorem are suitable to treat the system dynamics \cite{Yariv:1999,Poon:2004,Yariv:BOOK}, in contrast to the input-output formalism presented in the main text. 

\begin{figure}[b]
\includegraphics[width=0.6\textwidth]{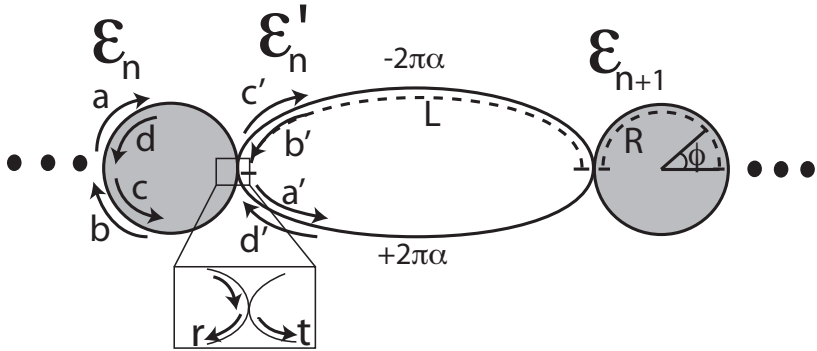}\caption{1D array of coupled resonators and waveguides\label{fig:1D-array}}

\end{figure}

We use the transfer matrix formalism for a 1D array of coupled resonators.
As shown in Fig.\ref{fig:1D-array}, the fields at the right edge
of the $n$-th resonator and the $n$-th waveguide are respectively characterized
by:

\begin{equation}
\mathcal{E}_{n}=\left(\begin{array}{c}
a_{n}\\
b_{n}\\
c_{n}\\
d_{n}\end{array}\right),\mathcal{E}'_{n}=\left(\begin{array}{c}
a_{n}'\\
b_{n}'\\
c_{n}'\\
d_{n}'\end{array}\right), \label{eq:fields} \end{equation}
such that the electric field inside the resonators is:

\begin{numcases}
{E_n(\rho,\phi)=f(\rho)\times}
a_n\exp\left[i\beta R(\frac{\pi-\phi}{\pi})\right]+d_n\exp\left[-i\beta R(\frac{\pi-\phi}{\pi})\right], & for $0<\phi<\pi$\\
b_n\exp\left[-i\beta R(\frac{\pi+\phi}{\pi})\right]+c_{n}\exp\left[i\beta R(\frac{\pi+\phi}{\pi})\right], & for $-\pi<\phi<0$\end{numcases}
where $\rho$ is the radial coordinate, $\phi$ is the azimuthal angle
relative to the counter-clockwise propagation direction and $f(\rho)$
is the radial dependence of the electric field.

We assume the resonators have length $2R$ and the total length of the
connecting waveguides is $2L$, as shown in Fig.\ref{fig:1D-array}.
 Therefore, transfer matrices corresponding to free propagation inside resonators and
waveguides are:

\begin{eqnarray}
M_{res} & (R)= & \left(\begin{array}{cccc}
e^{i\beta R} & 0 & 0 & 0\\
0 & e^{-i\beta R} & 0 & 0\\
0 & 0 & e^{i\beta R} & 0\\
0 & 0 & 0 & e^{-i\beta R}\end{array}\right)\\
M_{wg}(L) & = & \left(\begin{array}{cccc}
e^{i\beta L+i2\pi\alpha y} & 0 & 0 & 0\\
0 & e^{-i\beta L+i2\pi\alpha y} & 0 & 0\\
0 & 0 & e^{i\beta L-i2\pi\alpha y} & 0\\
0 & 0 & 0 & e^{-i\beta L-i2\pi\alpha y}\end{array}\right),\end{eqnarray}
respectively, where the wave number is $\beta=\omega/c$ and $\omega$
is the light frequency. We assume that the clockwise and counter-clockwise
modes, i.e., forward and backward fields, are decoupled from each
other, which sets the off-diagonal elements in $M_{wg}$ and $M_{res}$ to be zero. The regime where this coupling between forward and backward fields becomes important is discussed in the main text
and also in Sec.\ref{sec:back-scattering} of this supplemental information (SI). We further assume that the coupling regions are long enough (compared to the wavelength of light) so that the coupling between clockwise and counter-clockwise photons  (e.g. the coupling between $(a,b)$ and $(c',d')$ fields) can be ignored. In other words, a single coupling region couples only $ (a,b)$ to $(a',b')$ and $(c,d)$ to $(c',d')$.  Therefore, a loss-less coupling region can be characterized by two parameters  $t$ and $r$, which represent the amount of transmission into and reflection off of the adjacent element, respectively, as shown in Fig.\ref{fig:1D-array}.  Note that these coefficients obeys the conditions: $|t|^2+|r|^2=1$ and $r^*t+r t^*=0$. One can easily find the transfer matrix elements of such coupling regions for $ (a,b)$ and $(a',b')$, by setting $a=1,a'=t,b=r,b'=0$ and $a=0,a'=r,b=t,b'=1$, and similarly for coupling between $(c,d)$ and $(c',d')$. Therefore, the transfer matrix of a single coupling region which couples $ (a,b)$ to $(a',b')$ and $(c,d)$ to $(c',d')$ can be written as:
\begin{eqnarray}
M_{cpl} & = & \left(\begin{array}{cccc}
\frac{t^{2}-r^{2}}{t} & \frac{r}{t} & 0 & 0\\
-\frac{r}{t} & \frac{1}{t} & 0 & 0\\
0 & 0 & \frac{t^{2}-r^{2}}{t} & \frac{r}{t}\\
0 & 0 & -\frac{r}{t} & \frac{1}{t}\end{array}\right).\\
\end{eqnarray}
 
In this notation, the field amplitudes in
the right edge of the waveguide can be obtained from that of the adjacent
resonator: $\mathcal{E}_{n}'=M_{cpl}M_{res}\mathcal{E}_{n}$. 

The transfer matrix of the building block of such system ($\mathcal{E}_{n+1}=M\mathcal{E}_{n}$) is a product
of the above matrices $M=M_{cpl}M_{wg}M_{cpl}M_{res}$.
Using Bloch's theorem, the dispersion relation is written as:
\begin{equation}
\textrm{Det}[M-\exp(iK\Lambda)I]=0,\label{eq:bloch-theorem}\end{equation}
where $K$ is the Bloch quasi-momentum and $\Lambda$ is the unit spacing. First we solve the case where the phase imbalance between waveguide arms is zero
($\alpha=0$). We define the detuning from the resonator frequency
as $\Delta=\omega-\omega_{0}$. We consider a configuration where
resonators are separated by $L=(2p+1)\lambda_{0}/4$ where $p$ is a
non-zero integer and $\lambda_{0}$ is the resonant wavelength. In
this configuration, photon reflections off the resonators interfere
destructively with each other, and therefore, photons spends most
of their time in the resonators rather than the waveguides. We are
primarily interested in relatively high quality factor resonators.
Therefore, we evaluate the above expression to the first order in
$(1-|r|^{2})$, which is proportional to the inverse of the resonators
finesse $\mathcal{F}\simeq\pi/(1-r^{2}) \gg 1$. By assuming $r$ to be real
and $R=n\lambda_{0}$ for resonators length, Eq.(\ref{eq:bloch-theorem})
gives a dispersion relation: $\cos(K\Lambda)=(-1)^{p}\frac{4\pi n}{(1-r^{2})\omega_{0}}\Delta=(-1)^{p}\frac{\Delta}{2\kappa}$,
where $\kappa$ characterizes the cavity field decay rate from on
side of the resonator into the waveguides, i.e., the photon decay is $4\kappa=(1-r^{2})c/R$. This decay is the extrinsic desirable decay, in contrast to intrinsic loss  ($\kappa_{in}$) which we discuss in Sec.V.  Note that the sign of the tunneling
amplitude depends on the parity of $p$, which we assume to be odd ($p=2m+1$ in Fig.1a of the main text), so
that we get the conventional minus sign in the tight-binding Hamiltonian. 

Alternatively, one can also diagonalize the transfer matrix of a cell
and study its eigenvalues as propagating waves. Under the above assumption,
the diagonalized form of the transfer matrix of a unit cell will be
of the following form:

\begin{equation}
M_{diag}(0)=
-\left(\begin{array}{cccc}
\frac{\Delta}{2\kappa}+i\sqrt{1-(\frac{\Delta}{2\kappa})^{2}} & 0 & 0 & 0\\
0 & \frac{\Delta}{2\kappa}-i\sqrt{1-(\frac{\Delta}{2\kappa})^{2}} & 0 & 0\\
0 & 0 & \frac{\Delta}{2\kappa}+i\sqrt{1-(\frac{\Delta}{2\kappa})^{2}} & 0\\
0 & 0 & 0 & \frac{\Delta}{2\kappa}-i\sqrt{1-(\frac{\Delta}{2\kappa})^{2}}\end{array}\right)\end{equation}
We can infer: $-\left(\frac{\Delta}{2\kappa}\pm i\sqrt{1-(\frac{\Delta}{2\kappa})^{2}}\right)=e^{\pm ik\Lambda}$,
which is equivalent to the aforementioned dispersion relation. Again,
this shows that the propagating wave only exists when $|\frac{\Delta}{2\kappa}|\leq1$.
Note that in this situation, we have two modes with the same dispersion
relation, which corresponds to clockwise and counter-clockwise photons
inside the resonators. 

Furthermore, we consider the situation where there is a $2\pi\alpha$
phase imbalance between waveguide arms. In this case, the diagonalized
form of the transfer matrix takes the form:

\begin{equation}
M_{diag}(\alpha)=M_{diag}(0)
\left(\begin{array}{cccc}
e^{-2i\pi\alpha} & 0 & 0 & 0\\
0 & 
e^{-2i\pi\alpha} & 0 & 0\\
0 & 0 & 
e^{2i\pi\alpha} & 0\\
0 & 0 & 0 & 
e^{2i\pi\alpha}
\end{array}\right).\end{equation}
From Eq.(\ref{eq:bloch-theorem}) one can easily see that the above
transfer matrix corresponds to two dispersion relations: $\cos(K\Lambda\pm2\pi\alpha)=-\frac{\Delta}{2\kappa}=-\frac{\omega(K)-\omega_0}{2\kappa}$,
which is the same dispersion of the tight-binding Hamiltonian in 1D with a magnetic field:
\begin{eqnarray}
H_{0} & = & -\kappa \sum_{\sigma,x,y}\hat{a}_{\sigma x+1,y}^{\dagger}\hat{a}_{\sigma x,y}e^{\mp i2\pi\alpha y \sigma }+\hat{a}_{\sigma x,y}^{\dagger}\hat{a}_{\sigma x+1,y}e^{\pm i2\pi\alpha y \sigma }.
\end{eqnarray} 
In particular, the mode corresponding to (c,d) fields in the resonators (pseudo-spin
up) experiences a magnetic field ($+2\pi\alpha$) , while the mode
corresponding to (a,d) field (pseudo-spin down) experiences an opposite
magnetic field $(-2\pi\alpha)$. In a 2D arrangement, by choosing
these imbalances to be a function of rows $(\pm2\pi\alpha y)$, spin-up
(spin-down) photons experience a uniform magnetic field corresponding
to $+2\pi\alpha(-2\pi\alpha)$ flux per plaquette, respectively.

\section{Derivation of the Hamiltonian with scatterers\label{sec:spin-hall}}

In this section, we show that by implanting a scatterer in the system,
one can achieve a photonic version of different quantum spin Hall
Hamiltonians. In particular, the presence of a scatterer in each resonator
leads to an effective ``in-plane magnetic field'' and the presence
of scatterer in each connecting waveguides leads to a {}``spin-flip
hopping'' term. Below, we consider these effects.

\subsection{Spin-flip hopping}

Here, we study of the effect of scatterers in the connecting waveguides.
In particular, we position two scatterers symmetrically in the middle
of each waveguide, as shown in Fig.1(c) of the the main text. Therefore,
the transfer matrix of the scatterer which couples (a,d) and (b,c)
field together (as defined in Eq.\ref{eq:fields}), is given by:

\begin{equation}
M_{scatt}=\left(\begin{array}{cccc}
\frac{t_{s}^{2}-r_{s}^{2}}{t_{s}} & 0 & 0 & \frac{r_{s}}{t_{s}}\\
0 & \frac{1}{t_{s}} & -\frac{r_{s}}{t_{s}} & 0\\
0 & \frac{r_{s}}{t_{s}} & \frac{t_{s}^{2}-r_{s}^{2}}{t_{s}} & 0\\
-\frac{r_{s}}{t_{s}} & 0 & 0 & \frac{1}{t_{s}}\end{array}\right).\end{equation}

In this situation, the transfer matrix of the unit cell is given by: 

\begin{equation}
M=M_{res}\left(\frac{R}{2}\right).M_{res}\left(\frac{R}{2}\right).M_{cpl}.M_{wg}.\left (\frac{L}{2}\right).M_{scatt}.M_{wg}\left(\frac{L}{2}\right).\end{equation}
We assume the scatterer is weak so we can write the transmission/reflection
coefficients in the following form:

\begin{eqnarray}
r_{s} & = & i\sqrt{1-t_{s}^{2}}\\
t_{s} & = & 1-\epsilon^{2}/2,\epsilon\ll1\end{eqnarray}

In the absence of the scatterer, (a,b) and (c,d) modes couple to each
other independently, i.e. $(1,0,0,0)$ couples to $(0,1,0,0)$ and
$(0,0,1,0)$ couples to $(0,0,0,1)$. As we discussed, these decoupled
manifolds form an identical tight-biding Hamiltonian for two pseudo-spin
components which we call ($a_{\uparrow},a_{\downarrow}$). However,
the scatterer couples these two manifolds to each other. It turns
out that a proper basis in the presence of the scatterer is a Hadamard
rotated basis, i.e. $\frac{1}{\sqrt{2}}(1,0,\pm1,0)$ and $\frac{1}{\sqrt{2}}(0,1,0,\pm1)$.
This can be verified by using a degenerate perturbative method to
find eigenvalues of the transfer matrix. Therefore, one finds out
that the basis which diagonalizes M in the subspace of (a,c) or (b,d)
is the x-basis --i.e. $a_{\rightleftarrows}=\frac{1}{\sqrt{2}}(a_{\uparrow}\pm a_{\downarrow})$.
The eigenvalues of the system to the zeroth order in inverse of resonator
finesse $\propto(1-r^{2}$) are:

\begin{eqnarray}
\lambda_{\pm}^{\rightarrow} & =- & \frac{\left(2-\epsilon  \sqrt{4-\epsilon ^2}\right) \frac{\Delta}{2\kappa} \pm i \sqrt{\left(2-\epsilon ^2\right)^2+(4 \epsilon  \sqrt{4-\epsilon ^2} -4-4 \epsilon ^2+\epsilon ^4) (\frac{\Delta}{2\kappa})^2}}{2-\epsilon ^2}\\
\lambda{}_{\pm}^{\leftarrow} & =- & \frac{\left(2+\epsilon  \sqrt{4-\epsilon ^2}\right) \frac{\Delta}{2\kappa} \pm i \sqrt{\left(2-\epsilon ^2\right)^2+(-4 \epsilon  \sqrt{4-\epsilon ^2} -4-4 \epsilon ^2+\epsilon ^4) (\frac{\Delta}{2\kappa})^2}}{2-\epsilon ^2}.\end{eqnarray}

These eigenstates have magnitude equal to one. Since each pair of eigenvalues are complex conjugate of each other,
the four eigenvalues of system can be written in form of Bloch waves:\begin{eqnarray*}
\lambda_{\pm}^{\rightarrow} & = & e^{\pm iK\Lambda}\\
\lambda_{\pm}^{\leftarrow} & = & e^{\pm iK'\Lambda}\end{eqnarray*}
 This leads to two dispersion relations to the first order in $\epsilon$:

\begin{eqnarray*}
-\left(1\mp\epsilon\right)\frac{\Delta}{2\kappa} & = & \cos[K\Lambda]\end{eqnarray*}
 which can be simply written as:

\begin{eqnarray*}
\omega-\omega_{0} & = & -2\kappa(1\pm\epsilon)\cos[K\Lambda].\end{eqnarray*}

Now, we want to write this Hamiltonian in the second quantized form
using the original pseudo-spin components in z-basis ($a_{\uparrow},a_{\downarrow}$).
Note that the above dispersion relation is written in the Hadamard
rotated basis $(a_{\rightarrow},a_{\leftarrow})$ where $a_{\rightleftarrows}=\frac{1}{\sqrt{2}}(a_{\uparrow}\pm a_{\downarrow})$.
Therefore, we have:

\begin{equation}
H=-\sum_{<i,j>}(\begin{array}{cc}
a_{\rightarrow j}^{\dagger} & a_{\leftarrow j}^{\dagger}\end{array})\left(\begin{array}{cc}
\kappa(1+\epsilon) & 0\\
0 & \kappa(1-\epsilon)\end{array}\right)\left(\begin{array}{c}
a_{\rightarrow i}\\
a_{\leftarrow i}\end{array}\right).\end{equation}
The Hamiltonian in the original basis takes the conventional hopping
flip term:

\begin{equation}
H_{flip}=-\sum_{<i,j>}(\begin{array}{cc}
a_{\uparrow j}^{\dagger} & a_{\downarrow j}^{\dagger}\end{array})\left(\begin{array}{cc}
\kappa & \kappa\epsilon\\
\kappa\epsilon & \kappa\end{array}\right)\left(\begin{array}{c}
a_{\uparrow i}\\
a_{\downarrow i}\end{array}\right)\end{equation}
While the direct hopping amplitude is $\kappa$, the spin-flip hopping
amplitude is $\kappa\epsilon$. We note the above Hamiltonian is only
valid when the scatters are positioned symmetrically in the middle
of waveguides. If we move the scatterers symmetrically along the waveguides,
we arrive at the same form of Hamiltonian with a modified strength
of spin-flip term.

\subsection{In-plane magnetic field }

We now evaluate the effect of having
a scatterer in every resonator rather than the waveguides. As we show, this model results in an
in-plane magnetic field term in the Hamiltonian.

In this case the transfer matrix of the unit cell is: $M=M_{res}(R/2)M_{scatt}M_{res}(R/2)M_{cpl}M_{wg}(L/2)M_{wg}(L/2)$.
Writing again the transfer matrix in the x-basis, we perturbatively
find the eigenvalues to the zeroth order in $(1-r^{2})$ and the first
order in $(\epsilon)$:

\begin{eqnarray*}
\cos{K\Lambda}=\frac{1}{2}(\lambda_{+}^{\rightarrow}+\lambda_{-}^{\rightarrow})&\simeq&-\frac{\Delta}{2\kappa}-\frac{2\epsilon}{1-r^2}\\
\cos{K'\Lambda}=\frac{1}{2}(\lambda_{+}^{\leftarrow}+\lambda_{-}^{\leftarrow})&\simeq&-\frac{\Delta}{2\kappa}+\frac{2\epsilon}{1-r^2}
\end{eqnarray*}
 Therefore, the dispersion relations will be

\begin{eqnarray*}
\Delta & = & \mp\frac{4\epsilon\kappa}{(1-r^{2})}-2\kappa\cos[K\Lambda].\end{eqnarray*}
The corresponding Hamiltonian in the x-basis $(a_{\rightarrow},a_{\leftarrow})$
is:

\begin{equation}
H=\frac{4}{\pi}\sum_{i}(\begin{array}{cc}
a_{\rightarrow i}^{\dagger} & a_{\leftarrow i}^{\dagger}\end{array})\left(\begin{array}{cc}
-\epsilon\kappa\mathcal{F} & 0\\
0 & +\epsilon\kappa\mathcal{F}\end{array}\right)\left(\begin{array}{c}
a_{\rightarrow i}\\
a_{\leftarrow i}\end{array}\right)-\sum_{<i,j>}(\begin{array}{cc}
a_{\rightarrow j}^{\dagger} & a_{\leftarrow j}^{\dagger}\end{array})\left(\begin{array}{cc}
\kappa & 0\\
0 & \kappa\end{array}\right)\left(\begin{array}{c}
a_{\rightarrow i}\\
a_{\leftarrow i}\end{array}\right),\end{equation}
where the finesse of the resonators $\mathcal{F}\simeq\pi/(1-r^{2})$,
characterizes the average number of photon round trip before leaving
the resonator. The above Hamiltonian show that the scatterer inside
the resonator lifts degeneracy and splits the modes. The splitting
is proportional the strength of the scatterer and the number of photon
round trips inside the resonator. In the original z-basis $(a_{\uparrow},a_{\downarrow})$,
the Hamiltonian takes the form:

\begin{equation}
H_{mag}=-\frac{4\kappa\epsilon\mathcal{F}}{\pi}\sum_{i}(\begin{array}{cc}
a_{\uparrow i}^{\dagger} & a_{\downarrow i}^{\dagger}\end{array})\left(\begin{array}{cc}
0 & 1\\
1 & 0\end{array}\right)\left(\begin{array}{c}
a_{\uparrow i}\\
a_{\downarrow i}\end{array}\right)-\kappa \sum_{<i,j>}(\begin{array}{cc}
a_{\uparrow j}^{\dagger} & a_{\downarrow j}^{\dagger}\end{array})\left(\begin{array}{cc}
1 & 0\\
0 & 1\end{array}\right)\left(\begin{array}{c}
a_{\uparrow i}\\
a_{\downarrow i}\end{array}\right)\end{equation}
This means that the presence of the scatterer inside the resonators
result in an in-plane magnetic field. The amplitude of this in-plane
magnetic field is given by the strength of the scatterer and the number
of photon round trips inside the resonator. Similar to the spin-flip
term, the strength of this terms can modified by symmetrically moving
the scatterers long the resonators.

In the end, we note that the backscattering can also occur in the
waveguides. However, in contrast to the resonator case, where the
scattering was enhanced by the finesse of the resonators, the backscattering
effect in the waveguides is not enhanced due the off-resonant nature
of the waveguide loops.

\section{Input-Output formalism for probing the photonic system}

We now consider the effects of coupling ``probe'' waveguides to the system.
When
the system and probe waveguides are not coupled to each other the
Hamiltonian is given by $H_{0}+H_{probe}.$ The first term corresponds
to the system which we want to probe and the the second term corresponds
to free propagating light in the probing waveguide,  i.e., $H_{probe}=\sum_{k}\omega(k)|k\rangle\langle k|$ where $|k\rangle$ represents the state of one photon in the waveguide  with
wave number k. We note that since the system is linear  and we are interested in the linear response, we treat the system in the first quantized representation and do not use creation and annihilation operators. On the other hand, the resonator fields are coupled to the forward-
and backward-going field of the probing waveguides represented by
the following coupling to the tight-binding Hamiltonian:

\begin{eqnarray}
V & = & \frac{1}{\sqrt{\mathcal{L}}}\sum_{-\infty<k<+\infty}V_{k}^{in}|in\rangle \langle k|+V_{k}^{in}|k\rangle \langle in| \nonumber\\
 & + & \frac{1}{\sqrt{\mathcal{L}}}\sum_{-\infty<k<+\infty}V_{k}^{out}|out\rangle \langle k|+V_{k}^{out}|k\rangle \langle out|\end{eqnarray}
$|in\rangle$ ($|out\rangle$)  represents the photonic state of the resonator
which is coupled to the input- (or output-) waveguide. We assume
that resonators to be large enough so that each resonator mode couples
to single running wave in the probing waveguides. For instance, CCW
mode of the resonators is coupled to forward- (backward-) going light
field at the input (output) probing waveguides, respectively (Fig.1(b)
of the main text). Therefore, only positive (negative) momentum terms
have non-zero coefficient in input (output) Hamiltonian terms. $\mathcal{L}$ is
the length of the waveguides which enters as a normalization factor.

Now, we want to study the scattering
of an incoming plane wave $|k\rangle$ which is sent into the probing
waveguide. The effect of system-probe coupling is manifested
in the output light field ($|\psi\rangle$) which leaves the system
in the probing waveguides. Since the system is linear, one can also
look at this process from a single-photon picture: a photon comes
from the input waveguide, while the system is initially in the vacuum
state. Then, the photon interacts with system and eventually gets
scattered into different output channels. The photon leaves through
the probing waveguides and the system returns to original vacuum state. 

In order to study such transport effects, we follow the formalism
introduced in Refs.\cite{Fan:1999,Xu:2000} to investigate our system.
We start by Lippmann-Schwinger equation:\begin{equation}
|\psi\rangle=|k\rangle+\frac{1}{\omega_{k}-H_{sys}-H_{probe}+i\epsilon}V|\psi\rangle\end{equation}
which can be solved through iteration:

\begin{equation}
|\psi\rangle=\sum_{n=0}^{\infty}\left(\frac{1}{\omega_{k}-H_{sys}-H_{probe}+i\epsilon}V\right)^{n}|k\rangle.\end{equation}
In other words, the T-matrix which characterizes the scattering amplitudes
into different output channels can be written as:

\begin{equation}
\langle k'|T|k\rangle=\langle k'|\sum_{n=0}^{\infty}\left(\frac{1}{\omega_{k}-H_{sys}-H_{probe}+i\epsilon}V\right)^{n}|k\rangle\end{equation}
Since the total system is linear, each photon should be exchanged
an even number of times between the waveguide and the system to have
a non-zero contribution in the output field. Therefore, only even
powers in the above expression have non-zero contribution to the T-Matrix.
In other words,

\begin{eqnarray}
T_{k'k} & = & \delta_{k',k}+\frac{1}{\omega_{k}-\omega_{k'}+i\epsilon}\frac{1}{\mathcal{L}}\sum_{m,n}V_{k',m}G_{m,n}(\omega_{k})V_{n,k}\label{eq:T-matrix}\end{eqnarray}
where the Green's function is defined as: $G=G_{sys}^{0}+G_{sys}^{0}VG_{wg}^{0}VG_{sys}^{0}+\ldots$
and $G_{sys}^{0}(G_{wg}^{0})$ is the uncoupled Green's function of
the system (probe), respectively. Indices (m,n) sums over different
system eigenmodes, i.e., without the probing waveguides. The perturbative
series can be summed to yield the exact Green's function in the form:

\begin{equation}
G=\frac{G_{sys}^{0}}{1-\Sigma G_{sys}^{0}}\label{eq:greens-function}\end{equation}
where the self-energy of the coupled system is given by: $\Sigma_{m,n}=(VG_{wg}^{0}V)_{m,n}=\frac{1}{\mathcal{L}}\sum_{k}V_{m,k}\frac{1}{\omega-\omega_{k}+i\epsilon}V_{k,n}.$
We assume that the group velocity of the waveguide (in the absence
of the coupling to the system) is smooth around the resonators frequency.
In this limit, the interaction coefficient can be written as: 

\begin{equation}
V_{k}^{in}=V_{0} \,\,,\,\, V_{k}^{out}=
V_{0}
\end{equation}
where $V_{0}$ is the coupling coefficient at the wave vector corresponding
to the resonant frequency of the ring resonators (i.e. $V_{0}=V_{m=0,k=\omega_{0}/c}$).
We transform the momentum sum of the self-energy expression into an
integral. The real part of the integral corresponds to a constant
energy shift which can be absorbed in the energy definition. The imaginary
part of the self-energy corresponding to input channel is: $-i\frac{\nu}{2} |in \rangle\langle in |$, which shows that the field inside the input resonator will decay into the input waveguide at a rate $\nu=\frac{V_{0}^{2}}{v_{g}}$,
where $v_{g}$ is the group velocity at the resonator frequency. Similarly,
one can evaluate the output self-energy terms. The self-energy for the total system can be written as

\begin{equation}
 \Sigma=-i\frac{\nu}{2}|in\rangle \langle in|-i\frac{\nu}{2}|out\rangle \langle out|. 
 \end{equation}
Note that the Green's function in Eq.(\ref{eq:greens-function}),
can be also written as $G^{-1}(\omega)=\omega-H_{sys}-\Sigma$, which
can be represented as a $(N_{x}N_{y})\times(N_{x}N_{y})$ matrix and easily evaluated numerically.

Now, having the Green's function of the coupled system, we return
to the evaluation of the scattered field. We start by evaluating the
wavefunction of the transmitted field in the same input channel (as
shown in Fig.(1)b in the main text), which can be defined as $\psi_{t}(x)=\lim_{x\rightarrow+\infty}\langle x|\psi\rangle$
and the input field is $|k\rangle$ with $k>0$. One can write the
transmission coefficient in terms of the T-matrix: $\psi_{t}(x)=\frac{1}{\sqrt{\mathcal{L}}}\sum_{k'=-\infty}^{+\infty}e^{ik'x}T_{k'k}.$
By using the T-matrix expression in Eq.(\ref{eq:T-matrix}), and the
Green's function, we can formally write the wavefunction of the transmitted
field. The first term in the T-matrix gives the trivial unperturbed
wavefunction $\frac{1}{\sqrt{\mathcal{L}}}e^{ikx}$. The second term
can be written as:

\begin{eqnarray*}
 &  & \sum_{k'=-\infty}^{+\infty}\frac{e^{ik'x}}{\sqrt{\mathcal{L}}}\frac{1}{\omega_{k}-\omega_{k'}+i\epsilon}\frac{1}{\mathcal{L}}V_{0}^{2}G_{in,in}(\omega_{k}).\end{eqnarray*}
Again by transforming the momentum sum into an integral and performing
the contour integral, the pre-factor before the resonator summation
simply reduces to $-i\frac{V_{0}^{2}}{v_{g}}=-i\nu$. Therefore, the
transmission coefficient for a given frequency will become:

\begin{equation}
t(\omega)=1-i\nu G_{in,in}(\omega).\end{equation}
Similarly, one can obtain the other transport coefficients:

\begin{eqnarray*}
r'(\omega) & = & -i\nu G_{out,in}(\omega_{k}),\\
r(\omega) & = & -i\nu G_{in,in}(\omega_{k}),\\
t'(\omega) & = & -i\nu G_{out,in}(\omega_{k}).\end{eqnarray*}

\section{Back-scattering disorders \label{sec:back-scattering}}

The presence of scatterers in the system leads to mixing of pseudo-spin up and spin down, acting as a ``magnetic disorder''.
If this coupling is done in a controllable fashion, we arrive at the spin coupling Hamiltonians discussed earlier. However, in a experimental realization undesired backscatterings (e.g., due to surface roughness) lead to a similar coupling between spin-up and -down components. In most cases, this coupling is negligible.  This has been
the case for coupled resonators system \cite{Almeida:2004,Poon:2006,Liang:2006,Vlasov:2006,Xia:2007}. However, in very high-Q resonators, the mode coupling (e.g., due to surface roughness) is usually greater than the resonator decay rate. In this case, the resonator enhances of the backscattering effect, and a mode splitting occurs \cite{Kippenberg:2002,Borselli:2005}. We note that as long as the interest is in the linear regime (as in this article), very high-Q resonators are not required. Nevertheless for completeness, we study the backscattering effect by first characterizing it and then evaluating its effects. As we show below, the effect of backscattering is more pronounced in the resonators and can be modeled by a magnetic disorder Hamiltonian. Before showing the derivation of the magnetic disorder Hamiltonian, we present the result of such model.  In particular, we consider a case where at the resonator ($x',y'$),  a  scatterer is present and characterized by transmission (reflection) coefficient as $t_s\simeq1$ ($r_s=i\epsilon/\sqrt{2}$), respectively. The corresponding perturbation  will be: \begin{equation} H_{sc}=\frac{2\epsilon \kappa \mathcal{F}}{\pi} (\begin{array}{cc}
a_{\uparrow x',y'}^{\dagger} & a_{\downarrow x',y'}^{\dagger}\end{array})\left(\begin{array}{cc}
0 & e^{-i\phi}\\
e^{+i\phi} & 0\end{array}\right)\left(\begin{array}{c}
a_{\uparrow x',y'}\\
a_{\downarrow x',y'}\end{array}\right)\label{eq:mag-hamiltonian}\end{equation} where $\phi$ depends on the position of the scatterer inside the resonator.
The effect of such term can be understood in the following way: If we consider a forward-going spin-up wave, the disorder (to the first order in $\epsilon$) scatters the photon current into forward- and backward-going spin-down wave. Similar effects occur for other spin currents. This is in direct analogy to a ``magnetic disorder'' in the context of electronic QSHE, where electronic spin currents are not immune to magnetic impurities. Since our system is bosonic such impurities does not break the time reversal symmetry. However, their presence has a similar effect to magnetic impurities in fermionic QSHE. In particular, if we excite the system in  spin-up modes, due to backscattering, spin-down modes will be also populated. However, since spin-up and -down are coupled to different output channels (i.e., $(r',t)$ and $(r,t')$, respectively), the backscattering effect will be manifested in the reduction of signal in $r',t$ channels and some leakage in $r,t'$ channels.  Fig.4 in the main text shows different transport coefficients where we have assumed the backscattering is as strong as the dissipative loss. We observe that although the transport properties of magnetic bands states are affected by such magnetic impurities, edge state transport is robust, which is due the suppression of backscattering events. In particular, the scattering of a forward-going spin-up  into forward (backward) -going spin-down is inhibited due to energy (momentum) mismatch.

\begin{figure}
\includegraphics[width=0.6\textwidth]{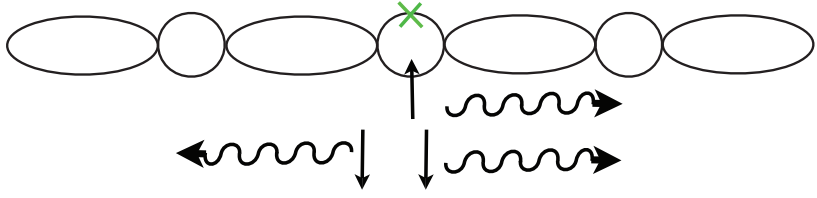}\caption{Scattering of a traveling spin-up photons from a resonator with a disorder}
\label{fig:backscattering-SI}

\end{figure}

Here we show that the effect of having an scatterer in a resonator
can be modeled by a simple perturbation in our Hamiltonian formalism. The scatterer
inside the resonator back-scatters the photon in the opposite mode,
which in our spin language means that the spin-up photons can be scattered
into spin down photons, vice verse. In particular, we consider the
following scenario: a spin-up forward propagating photon current,
in the presence of a back-scatterer in a resonator on its way, can
get back-scattered into forward- and backward-going spin-down current (Fig.\ref{fig:backscattering-SI}).
We first study this scenario in the physical system of a 1D array
of resonators by transfer matrix formalism to evaluate the forward-
and backward-going spin-down field amplitudes. Then, we study a model
Hamiltonian and evaluate the scattered field amplitudes. By comparing
the physical system and the model Hamiltonian, we argue that the physical
system can be faithfully modeled by our proposed Hamiltonian. 

First, we consider a 1D array of resonators and locate a disorder
inside a single resonator and evaluate the transport properties of
the a traveling spin wave off that disorder. In the absence of disorder
the system possesses four modes, corresponding to clockwise and counter-clockwise
modes of the resonators where each can propagate forwards or backwards.
Therefore, at each site the photonic field can be represented by a
4 dimensional vector. We work in the basis of this four modes: so
for example $(1,0,0,0)^{T}$ represents the forward spin-up photons
and similarly for the three other modes.

In this basis, as discussed in the Sec.I of this SI, the transfer
matrix of each cell is diagonalized in the following form:

\begin{equation}
M_{0}=-\left(\begin{array}{cccc}
e^{iK\Lambda} & 0 & 0 & 0\\
0 & e^{-iK\Lambda} & 0 & 0\\
0 & 0 & e^{iK\Lambda} & 0\\
0 & 0 & 0 & e^{-iK\Lambda}\end{array}\right)\end{equation}
where $\cos({K\Lambda})=-\frac{\Delta}{2\kappa}$ and $|\frac{\Delta}{2\kappa}|\leq1$.
Now, we consider that there is a cell with a disorder located at
site zero ($...M_{0}M_{0}M_{0}M_{scatt}M_{0}M_{0}...$). The propagation
of a free travelling spin wave can be deduced by application of the
transfer matrix $(M_{0}M_{scatt}M_{0}$). In particular, we are interested
in the reflected ($\mathcal{E}_{n}^{\leftarrow}$) and the transmitted
field $\mathcal{E}_{n+1}^{\rightarrow}$. However, these output field
are not readily available in the transfer matrix, and we have to use the
S-matrix. More precisely, $\left(\begin{array}{c}
\mathcal{E}_{n+1}^{\rightarrow}\\
\mathcal{E}_{n+1}^{\leftarrow}\end{array}\right)=M\left(\begin{array}{c}
\mathcal{E}_{n}^{\rightarrow}\\
\mathcal{E}_{n}^{\leftarrow}\end{array}\right)$ , while for S-matrix: $\left(\begin{array}{c}
\mathcal{E}_{n+1}^{\rightarrow}\\
\mathcal{E}_{n}^{\leftarrow}\end{array}\right)=S\left(\begin{array}{c}
\mathcal{E}_{n}^{\rightarrow}\\
\mathcal{E}_{n+1}^{\leftarrow}\end{array}\right)$. In our case, we deal with two forward- and two backward-modes, so
the M-matrix is characterized by 2x2 blocks as: $M=\left(\begin{array}{cc}
A & B\\
C & D\end{array}\right)$, therefore, the corresponding S-matrix will be given by $S=\left(\begin{array}{cc}
A-BD^{-1}C & BD^{-1}\\
-D^{-1}C & D^{-1}\end{array}\right)$.

We consider a situation where the disorder is in the middle of the
resonator in the upper arm. Therefore, the transfer matrix of two
cells with a disorder in between will be $M=M_{0}M_{scatt}M_{0}$.
The corresponding S-matrix to the first order in the disorder strength
($\epsilon$) and the zeroth order in the inverse of the finesse ($\mathcal{F}^{-1}=(1-r^{2})/\pi$),
will be:

\begin{equation}
S=e^{2iK\Lambda}\left(\begin{array}{cccc}
1 & 0 & 0 & 0\\
0 & 1 & 0 & 0\\
0 & 0 & 1 & 0\\
0 & 0 & 0 & 1\end{array}\right)-\frac{i\epsilon\mathcal{F}}{\pi\sin(K\Lambda)}e^{2iK\Lambda}\left(\begin{array}{cccc}
0 & 1 & 0 & 1\\
1 & 0 & 1 & 0\\
0 & 1 & 0 & 1\\
1 & 0 & 1 & 0\end{array}\right)\end{equation}
where the first part simply represents the free propagation over two
cells, and the second part represents the scattering into the other
modes due backscattering.  Note that as the scatterer is in the resonator,
its strength is enhanced by the finesses of the resonator. Now, we
consider a travelling spin-up wave which goes forward. For convenience,
we assume the phase of this wave is such that it is zero at the site
with the disorder. Therefore, our input will be: $v_{in}=\left(\begin{array}{c}
e^{-iK\Lambda}\\
0\\
0\\
0\end{array}\right).$ Since we are considering the propagation over two cells, in the absence
of the scatterer, the output field should be: $v_{output}^{\epsilon\rightarrow0}=S(\epsilon\rightarrow0)v_{in}=\left(\begin{array}{c}
e^{+iK\Lambda}\\
0\\
0\\
0\end{array}\right).$ However, in the presence of the scatterer, we will have: $v_{out}=e^{+iK\Lambda}\left(\begin{array}{c}
1\\
-\frac{i\epsilon\mathcal{F}}{\pi\sin(K\Lambda)}\\
0\\
-\frac{i\epsilon\mathcal{F}}{\pi\sin(K\Lambda)}\end{array}\right).$ The second (forth) row terms correspond to spin-down forward- (backward-)
going amplitudes. 

Now, we consider a tight-binding model with a disorder and show that
it has the same behavior as our photonic system. In particular, we
assume that the Hamiltonian of the system to be:\begin{equation}
H=-J\sum_{<i,j>}(\begin{array}{cc}
a_{\uparrow j}^{\dagger} & a_{\downarrow j}^{\dagger}\end{array})\left(\begin{array}{cc}
1 & 0\\
0 & 1\end{array}\right)\left(\begin{array}{c}
a_{\uparrow i}\\
a_{\downarrow i}\end{array}\right)-\epsilon'J(\begin{array}{cc}
a_{\uparrow0}^{\dagger} & a_{\downarrow0}^{\dagger}\end{array})\left(\begin{array}{cc}
0 & 1\\
1 & 0\end{array}\right)\left(\begin{array}{c}
a_{\uparrow0}\\
a_{\downarrow0}\end{array}\right)\label{eq:model-scattering}\end{equation}
where the first term corresponds to a simple spin-conserving hopping
Hamiltonian and the second terms corresponds to a magnetic disorder
at site zero which rotates the pseudo-spin along the x-axis at the specific
site. We assume the disorder to be weak ($\epsilon'\ll1$). Since
the system is linear, the dynamics of the system can be studied by
considering a single excitation. Therefore, a general state of the
system can be captured in a two-component wavefunction as a function
of sites: $|\Psi\rangle=\sum_{m}\psi_{\uparrow,\downarrow}(m)a_{\uparrow,\downarrow}^{\dagger}|0\rangle.$
In particular, since we are interested in the scattering of a forward-going
spin-up excitation off a disorder positioned at site m=0, the wavefunction
of the entire system can be described as:

\begin{numcases}
{\psi_{\uparrow}(m) =}
e^{iK\Lambda m}+r_{\uparrow}e^{-iK\Lambda m} & $m<0$\\
\psi_{\uparrow}(0) & $m=0$\\
t_{\uparrow}e^{+iK\Lambda m} & $m>0$\end{numcases}
\begin{numcases}
{\psi_{\downarrow}(m) =}
r_{\downarrow}e^{-iK\Lambda m} & $m<0$\\
\psi_{\downarrow}(0) & $m=0$\\
t_{\downarrow}e^{+iK\Lambda m} & $m>0$\end{numcases}
where $t_{\uparrow}$ ($t_{\downarrow}$), corresponds to the transmitted
amplitude of the spin-up (-down) component. Similarly, $r_{\uparrow}$
($r_{\downarrow}$), corresponds to the reflected amplitude of the
spin-up (-down) component. Now, using the above ansatz, we solve the
Schrodinger equation: $H|\Psi_{K}\rangle=E_{K}|\Psi_{K}\rangle$,
for $m=0,\pm1$ and $|m|>1$, which leads to seven independent equations: 

\begin{eqnarray}
E_{k} & = & -2J\cos(K\Lambda)\\
E_{k}\psi_{\uparrow}(0) & = & -J(e^{-iK\Lambda}+r_{\uparrow}e^{iK\Lambda})-Jt_{\uparrow}e^{+iK\Lambda}-\epsilon'J\psi_{\downarrow}(0)\\
E_{k}\psi_{\downarrow}(0) & =- & Jr_{\downarrow}e^{iK\Lambda}-Jt_{\downarrow}e^{+iK\Lambda}-\epsilon'J\psi_{\uparrow}(0)\\
E_{k}t_{\uparrow}e^{iK\Lambda} & = & -J\psi_{\uparrow}(0)-Jt_{\uparrow}e^{+i2K\Lambda}\\
E_{k}t_{\downarrow}e^{iK\Lambda} & = & -J\psi_{\downarrow}(0)-Jt_{\downarrow}e^{+i2K\Lambda}\\
E_{k}(e^{-iK\Lambda}+r_{\uparrow}e^{iK\Lambda}) & = & -J\psi_{\uparrow}(0)-J(e^{-2iK\Lambda}+r_{\uparrow}e^{+i2K\Lambda})\\
E_{k}r_{\downarrow}e^{iK\Lambda} & = & -J\psi_{\downarrow}(0)-Jr_{\downarrow}e^{+i2K\Lambda}.\end{eqnarray}
We can solve the above equations for an arbitrary value of $\epsilon'$.
However, since we are only interested in weak impurities, we present
the solutions to the first order in $\epsilon'$:

\begin{eqnarray}
t_{\uparrow}=1 & , & r_{\uparrow}=0\\
t_{\downarrow},r_{\downarrow} & = & \frac{i}{2}\frac{\epsilon'}{\sin(K\Lambda)}.\end{eqnarray}
These results mean that a forward-going spin-up wave scatters into
forward- and backward-going spin-down wave. This is identical to the
perturbative result of our physical system (i.e. resonators and waveguides).
In particular, if we set $J=\kappa$ and $\epsilon'=-2\epsilon\mathcal{F}/\pi$,
then the model Hamiltonian in Eq.\ref{eq:mag-hamiltonian} and Eq. \ref{eq:model-scattering} are equivalent to each other and they describe the scattering inside the resonator. We note the
the above derivations can be also reproduced for scattering of a spin-down
wave propagating in the forward or backward direction.

Moreover, we note that where the scatterer is not positioned in the
middle of the resonator the effective Hamiltonian is no longer an
in-plane magnetic field in the x-direction. For a general position,
the model Hamiltonian takes the form:

\begin{equation}
H=-J\sum_{<i,j>}(\begin{array}{cc}
a_{\uparrow j}^{\dagger} & a_{\downarrow j}^{\dagger}\end{array})\left(\begin{array}{cc}
1 & 0\\
0 & 1\end{array}\right)\left(\begin{array}{c}
a_{\uparrow i}\\
a_{\downarrow i}\end{array}\right)-\epsilon'J(\begin{array}{cc}
a_{\uparrow0}^{\dagger} & a_{\downarrow0}^{\dagger}\end{array})\left(\begin{array}{cc}
0 & e^{-i\phi}\\
e^{+i\phi} & 0\end{array}\right)\left(\begin{array}{c}
a_{\uparrow0}\\
a_{\downarrow0}\end{array}\right)\end{equation}
where the scattering terms corresponds to a magnetic field pointing
in a random direction in the system plane.

Furthermore, we note that one might be tempted to try an educated
guess based on the spin Hall Hamiltonian presented in Sec.\ref{sec:spin-hall},
without performing the above calculations. However, due the high symmetry
of the system considered in Sec.\ref{sec:spin-hall} (symmetric positioning
of the scatterers), we can not arrive at a rigorous approach to find
the exact form and the coefficients of Eq.(\ref{eq:model-scattering}).
In particular, varying the position of the scatterer in the single
disorder case leads to a magnetic field pointing in any direction
in the plane, while in the spin Hall model of Sec.\ref{sec:spin-hall},
symmetrically varying the position of the scatterers leads to only
a strength modification of the in-plane magnetic field.

\section{Effect of loss}

Here we discuss the effect of photonic loss which leads to leakage of photons out of the system.  The loss can occur both in resonators and
in the connecting waveguides. In most system, propagation
and bending losses dominate the light attenuation and one can ignore losses due to coupling regions \cite{Vlasov:2004,Xia:2007}. We characterize the propagation loss and the bending loss at $90^{o}$ turns by a single quantity $\mu$ which is the field attenuation in a roundtrip inside the resonator. By assuming moderate
quality factor for resonators and using transfer matrix formalism, one can easily show that the dispersion
relation of a 1D array of resonators takes the form: 

\begin{equation}
\omega(K)=\omega_{0}-2\kappa \cos (K\Lambda+2\pi\alpha y)-i\frac{2}{\pi}\mu\kappa\mathcal{F}.
\end{equation} 
This loss can be also represented by a non-hermitian term in the Hamiltonian:
\begin{equation}
H_{loss}=-i\kappa_{in}\hat{a}_{x}^{\dagger}\hat{a}_{x}
\end{equation} 
where the dissipation rate is given as the product of loss per roundtrip, number of roundtrips per decay time and the photon decay rate ($\kappa_{in}=\mu (\frac{\mathcal{F}}{2\pi}) (4\kappa)$).  

Fig.4(c) of the main text shows the reflection spectrum in the presence of such loss. In the edge state transfer band, the reflection is slightly attenuated due to the attenuation of the light field during the propagation around the system perimeter $(\simeq\exp(-4N_{x}\kappa_{in}))$. We note that the numerical results corroborate with this estimate.

\section{Experimental considerations}

We next consider experimental issues involving the realization and
detection of the magnetic photonic states. Although our formalism
is general and can be applied to various photonic systems, here, we
focus on silicon-on-insulator (SOI) technology where it has been shown
that up to 100 micro-ring resonators can be successfully coupled to
each other operating at 1.5$\mu m$ \cite{Xia:2007}. We consider
a 10x10 lattice and take the unit cell as 20$\mu m$.
Since for a typical ring of a few microns, bending losses ($\sim0.004$dB/turn) are typically
larger than propagation losses over a wavelength ($\sim3.5$dB/cm)\cite{Vlasov:2004,Xia:2007},
using waveguides with different lengths to connect the resonators
might be experimentally more accessible than coupling rings as in a Mach-Zender configuration \cite{Heebner:2004,Xia:2006}.
Using the experimental parameter in Ref. \cite{Xia:2007}, we find that the extrinsic decay is $\kappa=1 \rm{nm} (\frac{\omega_0}{\lambda_0})$, which is half of the bandwidth of the tight-binding model. The intrinsic loss is $\kappa_{in}\simeq 0.02 \rm{nm} (\frac{\omega_0}{\lambda_0})=0.02\kappa$ and the standard deviation of individual resonances is $U\simeq 0.4 \rm{nm} (\frac{\omega_0}{\lambda_0})=0.4\kappa$. In numerical simulations for Fig.4 and 5 of the main text, we use these parameters and assume that the backscattering strength is $\epsilon\mathcal{F}=0.1$.

\end{document}